\documentclass[acmtog]{acmart}
\acmSubmissionID{papers\_846s1}

\usepackage{makecell}
\usepackage{subcaption}

\usepackage[ruled]{algorithm2e} 

\SetAlFnt{\small}
\SetAlCapFnt{\small}
\SetAlCapNameFnt{\small}
\SetAlCapHSkip{0pt}

\copyrightyear{2025}
\acmYear{2025}
\makeatletter
\def\@ACM@copyright@check@cc{}
\makeatother
\setcopyright{cc}
\setcctype{by-nc-sa}
\acmConference[SIGGRAPH Conference Papers '25]{Special Interest Group on Computer Graphics and Interactive Techniques Conference Conference Papers }{August 10--14, 2025}{Vancouver, BC, Canada}
\acmBooktitle{Special Interest Group on Computer Graphics and Interactive Techniques Conference Conference Papers (SIGGRAPH Conference Papers '25), August 10--14, 2025, Vancouver, BC, Canada}
\acmDOI{10.1145/3721238.3730709}
\acmISBN{979-8-4007-1540-2/2025/08}

\begin{document}

\title{A Fast Parallel Median Filtering Algorithm Using Hierarchical Tiling}

\author{Louis Sugy}
\orcid{0009-0005-2134-3453}
\affiliation{%
 \institution{NVIDIA}
 \city{Munich}
 \country{Germany}}
\email{lsugy@nvidia.com}

\begin{abstract}
Median filtering is a non-linear smoothing technique widely used in digital image processing to remove noise while retaining sharp edges. It is particularly well suited to removing outliers (impulse noise) or granular artifacts (speckle noise). However, the high computational cost of median filtering can be prohibitive. Sorting-based algorithms excel with small kernels but scale poorly with increasing kernel diameter, in contrast to constant-time methods characterized by higher constant factors but better scalability, such as histogram-based approaches or the 2D wavelet matrix.

This paper introduces a novel algorithm, leveraging the separability of the sorting problem through hierarchical tiling to minimize redundant computations. We propose two variants: a data-oblivious selection network that can operate entirely within registers, and a data-aware version utilizing random-access memory. These achieve per-pixel complexities of $O(k \log(k))$ and $O(k)$, respectively, for a $k \times k$ kernel --- unprecedented for sorting-based methods. Our CUDA implementation is up to 5 times faster than the current state of the art on a modern GPU and is the fastest median filter in most cases for 8-, 16-, and 32-bit data types and kernels from $3 \times 3$ to $75 \times 75$.
\end{abstract}


%
%


\begin{teaserfigure}
\centering
\includegraphics[width=\textwidth]{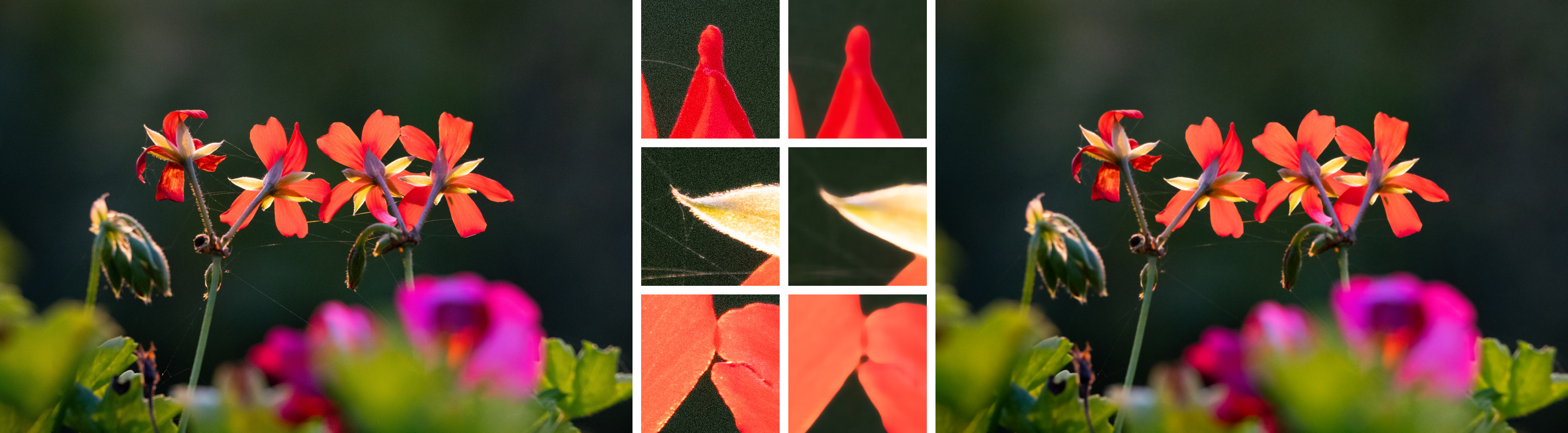}
\caption{A $17 \times 17$ median filter is applied to smooth a 30-megapixel photograph. The 8-bit red, green, and blue channels are filtered separately. Thanks to a computationally efficient and GPU-friendly algorithm, our method takes only 2.2 ms on an L40S GPU --- 3 times faster than the current state of the art.}
\label{fig:teaser}
\end{teaserfigure}

\maketitle

\section{Introduction}

The median filter \cite{tukey:1974} \cite{pratt:1975} is a fundamental tool in image processing. It replaces each pixel in an image with the median within a rectangular neighborhood, called a kernel. The only parameter, the kernel size, determines the trade-off between noise reduction and preservation of detail. This filter has several desirable properties, such as robustness to outliers and invariance to order-preserving transformations.



Median filtering has numerous applications as part of a broader image processing pipeline or model. It improves the accuracy of optical flow estimation by removing outliers in intermediate flow fields \cite{sun:2010}. It is commonly used as a pre-processing step for image segmentation to remove texture or noise without blurring edges, in particular in the context of medical imaging \cite{george:2017}. It can be used to denoise the output of stereo-matching \cite{zbontar:2016} and edge-detection algorithms \cite{topno:2019}. It is also used in photography and video editing software to create visually appealing images and artistic effects.

\subsection{The problem} \label{sec:intro-problem}

Unlike separable filters, the median filter cannot be expressed as the product of two 1-dimensional filters, posing a significant computational challenge. A naive approach processing each pixel independently, through the computation of a radix sort or a histogram, for instance, would at best have a complexity $O(k^2)$, traversing the entire $k \times k$ kernel for each pixel. Leveraging the overlap of kernels is required to avoid a quadratic complexity.

Aside from complexity, it is paramount to consider constant factors. Indeed, kernel diameters typically range from 3 to a few dozen. The most efficient median filtering methods fit into two broad categories. First, sorting-based methods of super-linear complexity. Second, constant-time methods with large constant factors.

Sorting-based methods are orders of magnitude faster for small kernels ($3 \times 3$ to $11 \times 11$) but experience a sharp performance drop as the kernel diameter increases. This limitation is particularly problematic for high-resolution image processing, where larger kernel sizes are often required. Histogram-based methods are limited to small data types such as 8 bits per channel, after which the histograms grow large, as do constant factors.

GPU architecture specifics are an integral part of designing an efficient algorithm. One of the most crucial considerations is which memory space to use for intermediate data. Registers offer the highest throughput, but they do not support dynamic indexing. Data can be backed by registers if accessed by a single thread and if the access pattern can be determined during compilation --- within the limit of available registers per thread. Another characteristic of the GPU programming paradigm is the \textit{Single Instruction, Multiple Threads} model (SIMT), in which threads of the same group, called warp, execute the same instruction in lockstep. Divergent control flow degrades performance because the divergent branches must be executed one after another.

For both reasons, a desirable property of the algorithm is to be data-oblivious, which means that the control flow and data accesses are independent of the input data. The opposite, an algorithm where the control flow or data accesses are determined dynamically from input data, is called data-aware.

\begin{figure}
\centering
\resizebox{0.9\columnwidth}{!}{%
\includegraphics[]{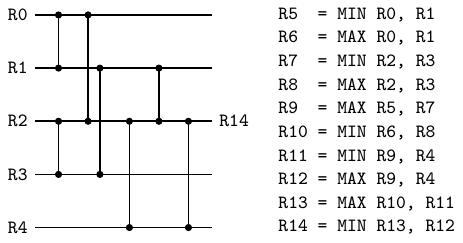}
}
\caption{A selection network that computes the median of 5 inputs. Knuth diagram \cite{knuth:1998} and the corresponding sequence of pseudo-assembly instructions.}
\label{fig:sort-network-sequence}
\end{figure}

The value of data-oblivious sorting algorithms was recognized early by Batcher \shortcite{batcher:1968}, who invented the odd-even and bitonic sorting networks. Sorting networks are sequences of compare-and-swap operations that sort a list in place. Those that only produce a subset of the sorted output, such as the minimum, maximum, or median values, are also called selection networks. Such networks can be statically generated and compiled into sequences of arithmetic instructions, as illustrated in Figure \ref{fig:sort-network-sequence}.

We jointly solve the two problems described above, of finding a method that: (a) scales better to larger kernel sizes (b) efficiently leverages the throughput of massively parallel graphics processors.

\subsection{Our contribution}

In this paper:

\begin{itemize}
  \item We introduce a novel median filtering algorithm using hierarchical tiling to leverage the separability of overlapping selection problems.
  \item We present a data-oblivious variant as a selection network with $O(k \log(k))$ complexity.
  \item We present a data-aware variant with $O(k)$ complexity.
  \item We describe GPU implementations of both variants.
  \item We benchmark our implementations against the state of the art, demonstrating significant performance improvements.
\end{itemize}

\section{Prior work}

\subsection{Histogram-based algorithms}

Huang et al. \shortcite{huang:1979} pioneered the study of computational aspects of median filtering. They observed that the kernels around two contiguous pixels share $k(k-1)$ values. A running histogram can be updated for each pixel by inserting $k$ values and removing $k$ values, to compute the medians with $O(k)$ per-pixel complexity.

This method was improved upon several times, first by Weiss \shortcite{weiss:2006} with a complexity $O(\log(k))$, then by Perreault and Hebert \shortcite{perreault:2007} with a complexity $O(1)$. They extended Huang's algorithm by maintaining running histograms for each image column, updating the main histogram associated with the sliding window in constant time. Green \shortcite{green:2018} described a parallel version of the constant-time median filter for GPUs; his implementation was the default in the GPU backend of the popular open-source library OpenCV \cite{opencv_library} until 2024.

However, the time complexity of these histogram-based methods hides high constant factors. The number of bins in the histograms is $\Theta(2^b)$, where $b$ is the number of bits used to represent a pixel. Traversing the histograms amounts to hundreds of cycles per pixel for 8-bit data types, and those methods are practically unusable for larger data types.

\subsection{Sorting-based algorithms} \label{sec:sorting-based}

Chakrabarti and Dhanani \shortcite{chakrabarti:1992} \shortcite{chakrabarti:1993} first proposed using sorting networks for median filtering. Sorting networks optimized for median selection were later used to implement graphics shaders for $3 \times 3$ and $5 \times 5$ median filters \cite{mcguire:2008}. 

These sorting networks were used in a perfectly parallel way: one was executed for each pixel independently. Using custom networks for small kernels, and constructs such as Batcher's odd-even merge sort, bitonic sort \shortcite{batcher:1968}, or Parberry's pairwise sorting network \shortcite{parberry:1992} for larger kernels, these implementations have an $O(k^2 \log(k)^2)$ complexity.

\begin{figure}
\centering
\resizebox{\columnwidth}{!}{%
\includegraphics[]{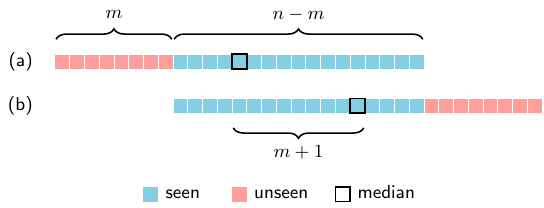}
}
\caption{Illustration of the principle of forgetfulness when selecting the median of $n$ elements and $m$ remain to be seen. We represent the array in ascending order from left to right, in two extreme cases: (a) the unseen values are the smallest $m$ values; (b) the unseen values are the largest $m$ values. The median is always contained in the middlemost $m+1$ values of $n-m$ seen so far.}
\label{fig:forgetful}
\end{figure}

Recently, there has been renewed interest in improving sorting-based algorithms due to increasing resolutions and adoption of 16- and 32-bit depths. These new algorithms leverage two principles: separability and forgetfulness. Separability is the idea of decomposing overlapping sorting problems to minimize redundant work. Forgetfulness is the principle that the median value can be found by iteratively including new values to a selection while discarding (forgetting) extrema. Extrema can be discarded when they are known to be greater, or smaller, than more than half of the inputs. As more values are included, the list of median candidates shrinks. When all inputs have been seen, the median is known. This principle is illustrated in Figure \ref{fig:forgetful}.

Perrot el al. \shortcite{perrot:2014} described the \textit{forgetful selection} algorithm: Starting from a subset of the kernel, they iteratively discard extrema and insert new elements into the selection. The size of the initial subset must be greater than $\lceil \frac{k^2}{2} \rceil + 1$, for the median element not to be discarded as an extremum through the process. They exploit the kernel overlap by computing two pixels per thread: the selection starts within the intersection of the two kernels, and only the final steps are performed separately.

Building on the idea of sharing work between contiguous pixels, Salvador et al. \shortcite{salvador:2018} extended the concept to two dimensions. For $k \ge 5$, they proposed using a forgetful selection with a $2 \times 2$ tile, sharing common work between all four pixels as well as pairs of pixels. Indeed, the intersection of the four overlapping kernels contains $(k - 1)^2$ inputs, out of the $k^2$ inputs in each kernel.

Adams \shortcite{adams:2021} generalized to larger tile sizes and suggested replacing the forgetful selection, which scales poorly, with a \textit{diagonal sorting network} followed by multiple merging steps. He demonstrated great performance breakthroughs, outperforming histogram-based methods for small and medium kernel sizes.

\subsection{2D Wavelet matrix}

Moroto and Umetani \shortcite{moroto:2022} proposed an extension of the wavelet matrix \cite{grossi:2003,claude:2015} for 2D arrays, that can be used for median filtering. After an initial construction step, this data structure supports constant-time queries for the median of any rectangular image region. Unlike histograms, the $O(b)$ complexity as a function of the pixel bit depth $b$ makes it suitable for high-precision data types. They showed that it outperforms histogram-based methods by an order of magnitude for 8-bit data on CUDA-enabled GPUs, and even sorting-based methods for large enough kernels ($k \ge 25$).

\section{Our algorithm} \label{sec:algorithm}

\subsection{Hierarchical tiling} \label{sec:hierarchical-tiling}

\begin{figure}
\centering
\resizebox{\columnwidth}{!}{%
\includegraphics[]{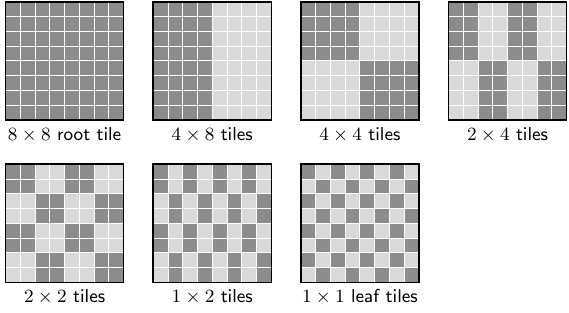}
}
\caption{Illustration of hierarchical tiling, starting from an $8 \times 8$ root tile. At each level, each tile is divided into two smaller tiles.}
\label{fig:hierarchical-tiling}
\end{figure}

Prior separable selection methods leverage tiling to mitigate redundant work. The image is partitioned into distinct tiles of size $t_w \times t_h$, and a single network computes $t_w t_h$ medians simultaneously.

However, the choice of tile size is a trade-off. If the tile is small, the early steps are not shared between as many pixels as possible. If the tile is large, the intersection between kernels is small, and the final non-shared steps dominate the cost. The intuition behind our new method is to separate the problem at multiple levels. That is, we want to share work between $n$ pixels, then $\frac{n}{2}$, $\frac{n}{4}$, and so on.

We propose the idea of hierarchical tiling, which can be represented as a binary tree: starting from a root tile, we recursively divide each tile into two smaller tiles. The recursion halts at $1 \times 1$ leaf tiles --- individual pixels.

The hierarchical tiling scheme that we use in the rest of this paper, illustrated in Figure \ref{fig:hierarchical-tiling}, splits square tiles horizontally, and non-square tiles on the longer side. Starting from a $\scriptstyle t_w^{(0)} \times \scriptstyle t_h^{(0)}$ root tile, where $\scriptstyle t_w^{(0)}$ and $\scriptstyle t_h^{(0)}$ are powers of two, the $2^i$ tiles at depth $i$ in the tree are of dimensions $\scriptstyle t_w^{(i)} \times \scriptstyle t_h^{(i)}$, recursively defined by:
\begin{equation*}%
\label{eqn:recursion}
\begin{cases}
t_w^{(i+1)} = \displaystyle \frac{t_w^{(i)}}{2},\quad t_h^{(i+1)} = t_h^{(i)} & \text{if $t_w^{(i)} \ge t_h^{(i)}$} \\
t_w^{(i+1)} = t_w^{(i)},\quad t_h^{(i+1)} = \displaystyle \frac{t_h^{(i)}}{2} & \text{else}
\end{cases}
\end{equation*}%

\subsection{Algorithm overview}

Following the principles of separability and forgetfulness described in Section \ref{sec:sorting-based}, the general idea of the algorithm is to include inputs progressively to a selection of median candidates while discarding extrema. Such a selection is attached to each tile at each level of the recursion, along with inputs that remain to be included. The key to performing the insertions efficiently is to keep this selection sorted, and remaining inputs partially sorted, as described below.

\begin{figure}
\centering
\resizebox{\columnwidth}{!}{%
\includegraphics[]{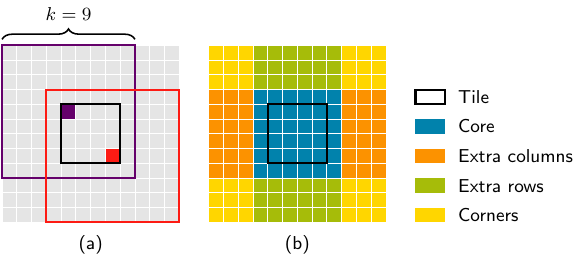}
}
\caption{Footprint of a $4 \times 4$ tile for a $9 \times 9$ kernel. (a) Outlines of the kernels for two pixels at opposite ends of the tile. (b) Partition of the tile's footprint into a core (the intersection of all $16$ kernels), extra columns, extra rows, and corners.}
\label{fig:tile-structure}
\end{figure}

For convenience, let us borrow the following terminology introduced by Adams \shortcite{adams:2021}, to refer to multiple classes of inputs in the context of a $t_w \times t_h$ tile and a $k_w \times k_h$ kernel, illustrated in Figure \ref{fig:tile-structure}:

\begin{description}
    \item[The footprint] is the union of the kernels associated with each pixel in the tile, of dimensions $(k_w + t_w - 1) \times (k_h + t_h - 1)$.
    \item[The core] is the intersection of the kernels. Its dimensions are $(k_w - t_w + 1) \times (k_h - t_h + 1)$.
    \item[Extra columns] are columns to the left and right of the core, with the same height as the core. There are $t_w - 1$ extra columns on each side.
    \item[Extra rows] are rows to the top and bottom of the core, with the same width as the core. There are $t_h - 1$ extra rows on each side.
    \item[Corners] are the other elements in the tile's footprint.
\end{description}

At each tile subdivision, the cores of the child tiles are supersets of the core of the parent tile, containing new extra rows or columns. For a $1 \times 1$ leaf tile at the end of the recursion, the core is exactly the kernel associated with the tile's unique pixel. The aforementioned list of median candidates is a sorted subset of the flattened core, minus excluded extrema, called the \textit{sorted core}.

The inputs that remain to be added to the selection are the extra columns, rows and corners. Each extra column and each extra row is kept sorted throughout the recursion, to minimize redundant sorting work between sub-tiles.

At any level of the recursion, the medians for all pixels in a tile are contained either in the sorted core, extra columns, extra rows or corners. At the end, for a leaf $1 \times 1$ tile, the extra rows, columns and corners are empty, and the sorted core of size 1 contains the median of the kernel associated with the tile's unique pixel.

The algorithm consists of two stages: the initialization that creates the data structures for the root tile, and the recursion that forks and updates the data structures from one parent tile to two child tiles.

\subsection{Initialization} \label{sec:algo-init}

The initialization comprises the following three operations with respect to the root tile:

\begin{itemize}
    \item \textbf{Sort the columns} of the core and the extra columns. The columns associated with horizontally contiguous tiles overlap, we can take advantage of this to share work between multiple tiles.
    \item \textbf{Sort the rows} of the core and the extra rows. Similarly, the rows associated with vertically contiguous tiles overlap.
    \item \textbf{Sort the core}. For this, we use a multi-way merge of either the sorted columns or the sorted rows.
\end{itemize}

These three initialization steps are also part of Adams' separable sorting networks \shortcite{adams:2021}, although the method used for sorting the core differs. Adams proposed a \textit{diagonal sorting network} composed of multiple sorting steps, where extrema are discarded at every step when possible: the core is first sorted column-wise, then row-wise, then diagonally, and finally a pairwise sorting network is applied to the remaining elements. However, if the majority of the core elements are retained, the last step is equivalent to a full sort. We found that using a multi-way merge of the sorted columns or rows was more efficient. In the data-oblivious variant of our algorithm, the multi-way merging network generally contains only half as many operations as the diagonal sorting network after pruning parts of the network that are unnecessary when discarding extrema.

\subsection{Recursion} \label{sec:algo-recursion}

\begin{figure}
\centering
\resizebox{0.97\columnwidth}{!}{%
\includegraphics[]{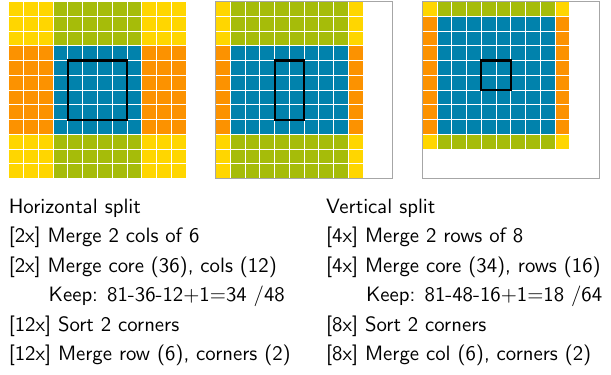}
}
\caption{First stages of the recursion for a $9 \times 9$ kernel using a $4 \times 4$ root tile, illustrated for the top-left child tile: the root tile is split horizontally into two $2 \times 4$ tiles, then again vertically into four $2 \times 2$ tiles.}
\label{fig:split}
\end{figure}

As described in Section \ref{sec:hierarchical-tiling}, the recursion consists in successive subdivisions from a root tile to $1 \times 1$ leaf tiles. Each split operation forks and updates the data structures associated with a parent tile to two child tiles, as follows:

\begin{description}
\item[Horizontal split] of a $\scriptstyle t_w^{(i)} \times \scriptstyle t_h^{(i)}$ tile. For each child tile:
\begin{itemize}
    \item $\scriptstyle t_w^{(i)} /2$ extra columns are merged into a single sorted list and then with the sorted core of the parent tile, to form the sorted core of the child tile. Extrema are discarded.
    \item For each extra row, $\scriptstyle t_w^{(i)} /2$ corners are sorted and merged with the extra row, to form the extra rows of the child tile.
\end{itemize}
\item[Vertical split] of a $\scriptstyle t_w^{(i)} \times \scriptstyle t_h^{(i)}$ tile. For each child tile:
\begin{itemize}
    \item $\scriptstyle t_h^{(i)} /2$ extra rows are merged into a single sorted list and then with the sorted core of the parent tile, to form the sorted core of the child tile. Extrema are discarded.
    \item For each extra column, $\scriptstyle t_h^{(i)} /2$ corners are sorted and merged with the extra col\-umn, to form the extra col\-umns of the child tile.
\end{itemize}
\end{description}

A horizontal and a vertical split are illustrated in Figure \ref{fig:split}. Pseudo-code is also provided in the supplemental material.

\section{Data-oblivious selection network} \label{sec:data-oblivious}

As explained in Section \ref{sec:intro-problem}, data-oblivious control flow and memory access patterns enable using registers, the fastest kind of memory, for most of the intermediate data. This also avoids a lot of arithmetic operations related to index calculations and control flow.

\subsection{Networks}

For the broader algorithm to be data-oblivious, the sorting and merging networks used in implementing its parts must satisfy the same constraint.

Our algorithm requires efficient networks, optimized for selecting a subset of the output (extrema are discarded at every step). For each type of network and problem dimension, we selected a few candidate networks, applied optimizations, and selected the one with the fewest arithmetic operations. The base networks that generally performed best for each task are the following:

\begin{description}
    \item[Sorting] For sizes up to 64, the best networks found with evolutionary methods \cite{dobbelaere:2024}. Above that, Parberry's pairwise sort \shortcite{parberry:1992}.
    \item[Merging ] A generalization of Batcher's odd-even merging network \shortcite{batcher:1968} to arbitrary input sizes.
    \item[Multi-way merging] Lee and Batcher's multi-way merging network \shortcite{lee:1995}.
\end{description}

\subsection{Complexity} \label{sec:oblivious-proof}

Let $k$ be an odd integer, $k \ge 3$. The per-pixel complexity of a $k \times k$ median filter using our selection network is $O(k \log(k))$.

\begin{proof}
A complete proof is provided in the supplemental material. Here, we only describe its high-level outline.

Let us define the following heuristic to choose a root tile:
\begin{equation*}%
    t_w^{(0)} = t_h^{(0)} = t(k) = 2^{\left\lfloor \log_2(k) \right\rfloor - 1}
\end{equation*}%

This heuristic guarantees that the root tile size grows linearly with the kernel size:
\begin{equation*}%
    \frac{k}{4} < t(k) < \frac{k}{2} \quad \text{so}~t(k) = \Theta(k) \label{eqn:rel-t-k}\\
\end{equation*}%

Based on Batcher and Lee's analysis \shortcite{batcher:1968} \shortcite{lee:1995}, we know upper bounds on the complexity of the following networks\footnote{We choose to ignore data-oblivious sorting networks with $O(n \log(n))$ complexity that are not practical to use due to large constant factors, such as Goodrich's zig-zag sorting network \shortcite{goodrich:2014}.}:

\begin{description}
    \item[Sorting network] for a list of size $n$: $O(n \log(n)^2)$.
    \item[Merging network] for two sorted lists of sizes $p$ and $q$: \\ $O((p+q) \log(p+q))$.
    \item[Multi-way merging network] for $k$ sorted lists of size $n$: \\ $O(k n \log(k) \log(n))$.
\end{description}

We can use this to show that the cost of the initialization stage is $O\left( k^2 \log(k)^2 \right)$ per root tile, and the total cost of the recursion is $O\left( t(k)^2 k \log(k) \right)$. Thus, the per-pixel cost is $O\left( k \log(k) \right)$.
\end{proof}

\subsection{CUDA implementation} \label{sec:oblivious-cuda-impl}

We map each root tile to a CUDA thread so a single thread can execute the full recursion. This avoids the overhead of synchronization and data exchanges between threads through shared or global memory. The only exception is the collaborative column sort in the initialization stage, as discussed below.

We leverage template metaprogramming, constant expressions, and loop unrolling to ensure that sorting and merging networks get compiled into sequences of min-max instructions, with all intermediate data stored in registers. This requires the CUDA functions to be compiled for each combination of parameters (kernel and tile dimensions).

To efficiently load the image data from global memory and avoid redundant work in the column sort, each thread block processes contiguous tiles, horizontally and vertically. For instance, a block of 512 threads can process 128 columns and 4 rows of tiles --- the most efficient configuration is chosen per kernel size and data type. We use shared memory to exchange data between threads in the same block. The CUDA function comprises the following stages:

\begin{enumerate}
    \item Load a region of the input image collaboratively from global memory into shared memory. The region is the union of the footprints of all the tiles assigned to the thread block. Synchronize the thread block.
    \item Sort the columns (core and extra columns) collaboratively per block row. The columns are mapped according to a round-robin pattern, loaded from shared memory to registers, sorted, and finally exchanged using shared memory.
    \item Load the footprint of the thread's assigned tile from shared memory into registers.
    \item Run the remaining stages of the hierarchical separable sorting network (the recursion).
    \item Store the medians of all the pixels in the tile into the output image in global memory.
\end{enumerate}

Although this approach is very efficient for small kernels, its applicability is limited by the number of available registers. First, there is a limit on the number of registers per thread (255 on all CUDA micro-architectures since Maxwell). Second, the size of the register file in each Streaming Multiprocessor (SM) is also limited, restricting the number of threads that can be scheduled simultaneously on each SM. When using 255 registers per thread, only 8 warps can be scheduled on each SM --- GPUs can otherwise fit a maximum of 32 to 64 warps per SM. The consequences of using too many registers are twofold:

\begin{description}
    \item[Limited occupancy] Few warps per SM. This can hinder the scheduler's ability to hide memory latencies by switching between warps.
    \item[Memory spilling] If there are not enough registers to store a thread's local data, it spills to local memory, a slower memory space that is backed by DRAM and the cache hierarchy.
\end{description}

The number of registers required to store the state associated with a tile grows with the kernel size, and the two effects above cause a decline in the performance of this implementation after $15 \times 15$ kernels, as seen in Figure \ref{fig:benchmark}.

\section{Multi-pass data-aware algorithm} \label{sec:data-aware}

The data-aware variant complements the data-oblivious one, achieving higher performance for large kernel sizes. For such sizes, storing all intermediate data in registers is impossible. Dropping the constraint of data obliviousness, we can use more efficient sorting and merging algorithms.

\subsection{Sorting and merging}

Merging two sorted lists sequentially in linear time is straightforward. Begin by initializing two iterators at the start of each list. Select the smaller value between the two, move the corresponding iterator forward, and repeat this process until all elements are merged. It is also possible to merge lists in parallel with a linear cost using the \textit{merge path} algorithm \cite{odeh:2012}: the output list is partitioned between threads, each of which uses a binary search to find the corresponding starting indices in both input lists.

Merging more than two lists efficiently on GPUs is a more challenging problem. It is possible to do with linear complexity, but it is preferable to trade high constant factors for a logarithmic factor. Indeed, an efficient algorithm on GPUs consists in merging lists pairwise following a binary reduction pattern --- the number of lists is halved at each iteration. This is parallelized in two ways: threads can work on each pair of lists to merge in parallel, and multiple threads can work on one merge operation using the merge path algorithm.

Various algorithms are suitable for parallel sorting, most notably the radix sort of linear complexity, a staple in the high-performance CUB library \cite{stehle:2017, adinets:2022}. In practice, when sorting many small arrays in parallel, the most efficient method is distributing the sorts between threads and using sorting networks.

\subsection{Complexity} \label{sec:aware-complexity}

Let $k$ be an odd integer, $k \ge 3$. The per-pixel complexity of a $k \times k$ median filter using the hierarchical-tiling median filtering algorithm is $O(k)$.

\begin{proof}
A complete proof is included in the supplemental material. We follow the same steps as Section \ref{sec:oblivious-proof}, updated to consider data-aware merging algorithms. Although we know linear algorithms for such operations, the complexities of the algorithms that we use in our implementation are sufficient assumptions to prove the linear complexity of the broader algorithm:

\begin{description}
    \item[Merging] two sorted lists of sizes $p$ and $q$: $O(p+q)$.
    \item[Multi-way merging] $k$ sorted lists of size $n$: $O(kn \log(k))$.
\end{description}

The cost of the initialization stage is still $O\left( k^2 \log(k)^2 \right)$ per root tile, but the total cost of the recursion is now $O \left( t(k)^2 k \right)$. Thus, the per-pixel cost is $O\left( k \right)$.

\end{proof}

\subsection{CUDA implementation}

CUDA exposes several memory spaces to the programmer, characterized by different scopes, access speeds, and sizes. Reading from shared memory is generally more than ten times faster than streaming data from DRAM. However, much like registers, the amount of shared memory per thread block is limited and impacts the number of thread blocks that can be active on the same SM. It is desirable to use shared memory for most of the intermediate data but for large kernels, the space required to store all the data associated with one root tile does not fit in shared memory.

The number of threads per merging problem is another very important consideration. When using too many threads, the binary search dominates the cost, as the number of elements per thread is of the same order as the logarithm of the number of elements in the array. But we cannot use too few threads per problem either, due to the aforementioned memory constraints. Efficiently mapping work to threads is particularly challenging in the context of a recursion involving many merging problems in different numbers and dimensions.

\begin{figure}
\centering
\resizebox{\columnwidth}{!}{%
\includegraphics[]{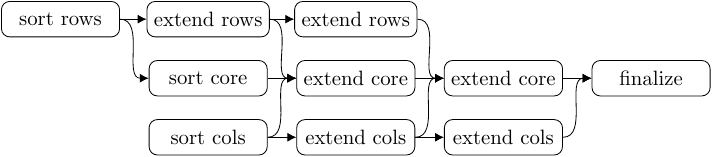}
}
\caption{Compute graph of the multi-pass algorithm, using an $8 \times 8$ root tile. Each box is a CUDA kernel, and each arrow represents an execution dependency, either implicit through streams or explicit, using events.}
\label{fig:multi-pass}
\end{figure}

This motivated a multi-pass approach, breaking the recursion into multiple stages. Each one processes all tiles, saving the intermediate state to global memory. To save DRAM bandwidth, we combined two recursion levels per pass: one horizontal and one vertical split, halving the tile width and height each time. This implementation is based on the following CUDA kernels:

\begin{description}
    \item[Row sort] Sorts core rows and extra rows.
    \item[Column sort] Sorts core columns and extra columns.
    \item[Core sort] Sorts the core by merging sorted rows.
    \item[Row extension] Reads corners from the input image and inserts them into the sorted rows (for horizontal split).
    \item[Column extension] Reads corners from the input image and inserts them into the sorted columns (for vertical split).
    \item[Core extension] Subdivides one tile into four smaller tiles, extending the core with extra rows and extra columns on each side (fused horizontal and vertical split).
    \item[Finalization] Similar to the core extension but optimized to avoid the final row extension, and writes the median to the output image.
\end{description}

The first three are used once in the initialization from Section \ref{sec:algo-init}, and the following three are repeated at each recursion step to implement the splits described in Section \ref{sec:algo-recursion}. Operations on rows, columns, and the core can be overlapped in three different CUDA streams, using events to manage execution dependencies between streams. The compute graph is represented in Figure \ref{fig:multi-pass}.

We added a pre-processing step to transpose the image, creating a column-major copy of the row-major input. The column-major image is used in operations on rows and the row-major image in operations on columns. This helps to achieve memory access coalescing and higher cache hit rates.

In contrast with the data-oblivious version, redundant operations on rows and columns are avoided. Indeed, the footprints associated with contiguous tiles largely overlap, as the kernel size is typically around twice the tile size. In this implementation, we avoid redundant work by sharing intermediate data between multiple tiles at every level of the recursion. Sorted rows are shared between vertically contiguous tiles, and sorted columns between horizontally contiguous tiles.

\section{Results} \label{sec:benchmark}

\begin{figure*}
\centering
\resizebox{\textwidth}{!}{%
\includegraphics[]{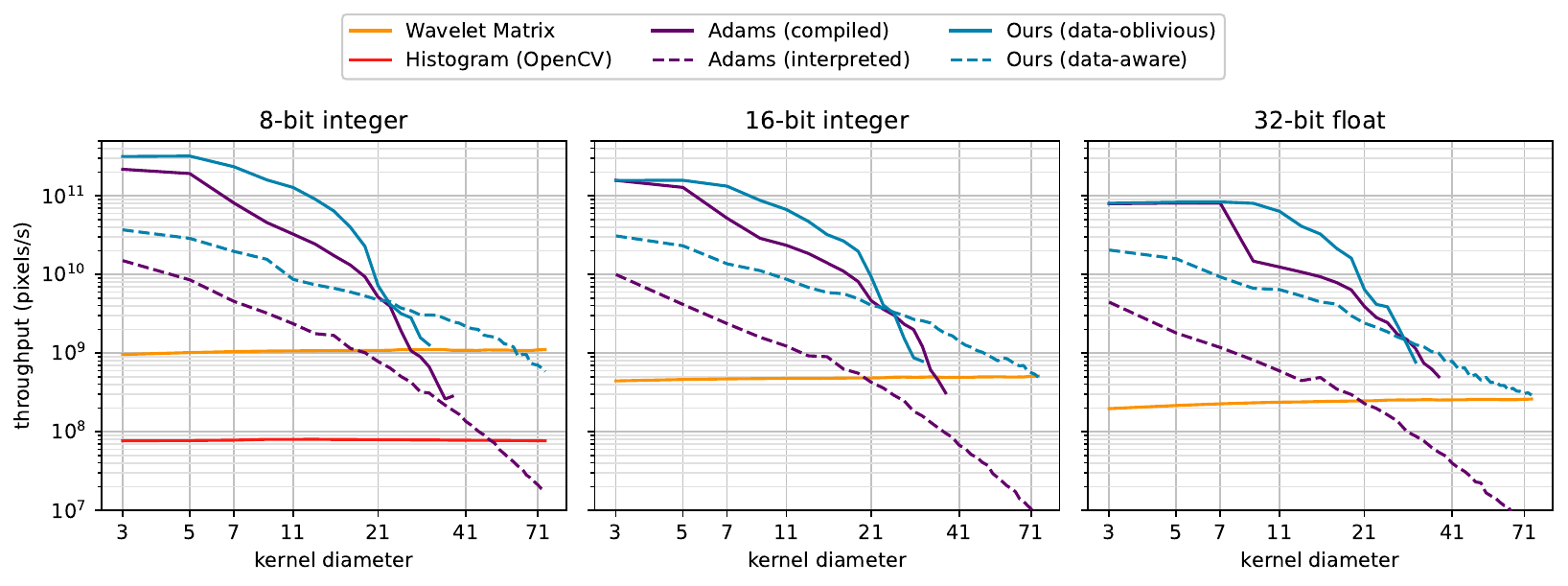}
}
\caption{Pixel throughput (higher is better) of various median filtering methods on an NVIDIA L40S GPU. The two implementations of our algorithm combined achieve the highest performance for a wide range of kernel sizes, until the constant-time 2D wavelet matrix method eventually becomes faster.}
\label{fig:benchmark}
\end{figure*}

To evaluate our algorithm, we benchmark against the best available implementations in each category of methods. We cover kernel sizes from $3 \times 3$ to $75 \times 75$.

Our setup is an NVIDIA L40S GPU, with 142 Ada Streaming Multiprocessors. We compute the pixel throughput from the median run time over multiple iterations with different 30-megapixel images. Data transfers between host and GPU memory are not measured.

We compare the following GPU implementations:

\begin{itemize}
    \item OpenCV's histogram-based median filter \cite{green:2018} (note: recent versions of OpenCV use the 2D wavelet matrix method by default; this can be disabled by editing the code).
    \item Moroto and Umetani's median filter using a 2D wavelet matrix \shortcite{moroto:2022}: we used the benchmark published on their project page.
    \item Adams' separable sorting networks \shortcite{adams:2021}: we used the benchmark published by Adams in the ACM digital library.
    \item Our hierarchical tiling algorithm using the two implementations described in Section \ref{sec:data-oblivious} and Section \ref{sec:data-aware}.
\end{itemize}

The benchmark results are shown in Figure \ref{fig:benchmark}.

These numbers show that sorting-based methods are the best performers for small kernels. For the smallest kernel sizes, the bottleneck is loading and storing the image from and to DRAM. Then, the data-oblivious variant of our method outperforms Adams' compiled (static) implementation, until both see a sharp performance drop due to the register pressure and spilling to local memory.

Around kernel sizes $23 \times 23$ (8 bits) to $29 \times 29$ (32 bits), the data-aware variant of our method becomes the fastest due to its linear complexity, as discussed in Section \ref{sec:aware-complexity}, widening the gap with Adams' -- for $75 \times 75$, it is 50 times faster.

Eventually, due to the constant complexity of the 2D wavelet matrix method, the latter becomes faster for the largest kernel sizes ($61 \times 61$ for 8-bit, $75 \times 75$ for 16-bit integers).

\section{Conclusion}

This paper introduced a new parallel median filtering algorithm, characterized by low complexity and constant factors. We demonstrated how to implement it on GPUs efficiently, outperforming the fastest existing implementations throughout most of the range of kernel sizes from $3 \times 3$ to $75 \times 75$.

Suitable for high-resolution images, large kernels, and high-pre\-ci\-sion data types, our algorithm enables the use of a filter that has long been thought too computationally expensive for practical use, especially in real-time systems.

\subsection{Limitations}

One limitation of our method is the compilation time (around 15 minutes) and the binary size (around 40 MB) when supporting all the kernel sizes and data types used in the benchmark in Section \ref{sec:benchmark}. The detailed compilation times per kernel size and per compilation phase can be found in the supplemental material.

Another limitation is the data-aware version's heavy memory requirement, exceeding the size of the input image by up to two orders of magnitude. This can be worked around by applying the filter iteratively to rectangular slices of the image to meet an arbitrary memory budget. The memory requirements are also detailed in the supplemental material.

\subsection{Future work}

One of the pain points affecting the performance of our method is the lack of an efficient parallel algorithm for multi-way merging, currently implemented in the data-aware version with successive two-way merges. This could be improved in future work.

It would also be useful to extend the idea of hierarchical tiling to 3D. Indeed, 3D median filters are often used in medical imaging. Efficient sorting-based filters have been proposed for small kernels \cite{crookes:2006}, but hierarchical tiling has the potential to enable larger kernel sizes.

\begin{acks}
Thanks to Tamás Béla Fehér for his support and advice. Thanks to colleagues for their valuable feedback, especially Dawid Pająk, whose comments were instrumental in giving the paper its present shape. Thanks to the anonymous reviewers for their thorough reviews and suggestions.
\end{acks}

\bibliographystyle{ACM-Reference-Format}
\bibliography{references}

\end{document}